# User-based collaborative filtering approach for content recommendation in OpenCourseWare platforms

Nikola Tomašević, Dejan Paunović, Sanja Vraneš
University of Belgrade, Institute Mihajlo Pupin, Belgrade, Serbia
nikola.tomasevic@pupin.rs, dejan.paunovic@pupin.rs, sanja.vranes@pupin.rs

*Abstract* — **A content recommender system or a recommendation system represents a subclass of information filtering systems which seeks to predict the user preferences, i.e. the content that would be most likely positively "rated" by the user. Nowadays, the recommender systems of OpenCourseWare (OCW) platforms typically generate a list of recommendations in one of two ways, i.e. through the content-based filtering, or user-based collaborative filtering (CF). In this paper, the conceptual idea of the content recommendation module was provided, which is capable of proposing the related decks (presentations, educational material, etc.) to the user having in mind past user activities, preferences, type and content similarity, etc. It particularly analyses suitable techniques for implementation of the user-based CF approach and user-related features that are relevant for the content evaluation. The proposed approach also envisages a hybrid recommendation system as a combination of user-based and content-based approaches in order to provide a holistic and efficient solution for content recommendation. Finally, for evaluation and testing purposes, a designated content recommendation module was implemented as part of the SlideWiki authoring OCW platform.**

I. INTRODUCTION

As defined by Francesco et al. [1], a content recommender system or a recommendation system represents a subclass of information filtering systems which seeks to predict the user preferences, i.e. the content that would be most likely positively "rated" by the user. Nowadays, recommender systems typically generate a list of recommendations in one of two ways (Jafarkarimi et al. [2]), through:

- Content-based filtering, or
- User-based collaborative filtering.

The goal of this paper is to provide a conceptual idea of the content recommendation module as part of the OpenCourseWare (OCW) platform, capable of proposing the related decks (presentations, educational material, etc.) to the user having in mind past user activities, preferences, type and content similarity, etc. To provide a holistic and efficient solution for content recommendation, the proposed solution envisages combination of user-based and content based approaches under the umbrella of hybrid recommendation system. However, the main focus of this paper is providing the underlying idea for the user-based collaborative filtering (CF) approach, which is elaborated in more detail in the following.

As part of the proposed hybrid content recommendation system, user-based CF will be implemented in addition to the content-based recommendation. The goal of user-based CF approach (John et al. [3]) will be to discover the users that share similar preferences and undertake similar activity patterns in order to make the content recommendation. Recommendations of the related content (i.e. decks) to the OCW user will be performed based on relevant user-related features. The user-related features will include user activities (duration and number of deck visits, deck creation/editing, participation to related discussion, deck commenting, etc.) and user preferences (likes and ratings of decks, etc.). Moreover, to make the recommendation process context-aware, user demographic & contextual data (e.g. age, skills, date and time, location, etc.) could be also taken into account. Leveraged upon relevant user-related features (e.g. past activities and preferences), the proposed solution would evaluate the similarity with other users by comparison with their activities and preferences. Based on similarity search, the most similar users (to the user whom the prediction is for) will be taken into account for extraction of the OCW content to be recommended. Recommendation will be made based on the content that is associated to the most similar users, but is not associated to the analysed user (who did not visited or rated it). In this way, by comparing the past user activities and preferences, the OCW platform will be capable to derive the content for recommendation.

In addition to the user-based CF approach, the proposed hybrid recommendation approach would also provide recommendations based on content "popularity". The popularity of the content will be evaluated based on visits, edits, comments or number of likes and average rates given by the users (either total or during the last week/month). In this way, these distinct but complementary approaches could be combined to provide more reliable recommendations leveraged upon both user-based information and content "popularity". Finally, for evaluation and testing purposes, the proposed user-based CF approach was implemented as a designated content recommendation module as part of the SlideWiki authoring OCW platform. SlideWiki platform is being developed as a part of the H2020 EU research and innovation project SlideWiki (Grant Agreement No 688095) aiming to create a large scale accessible learning and teaching platform, using educational technology, skill recognition and global collaboration and as such represents suitable but challenging testbed for content recommendation systems.

The remainder of this paper is organized as follows. The proposed approach and underlying methodology leveraged by k-NN technique is described in Section 2. Section 3 elaborates user-related features which are relevant for user-based CF approach. Implementation of content recommendation module as part of the SlideWiki authoring OCW platform is described in Section 4. Finally, concluding remarks and discussion are given in Section 5.

## II. PROPOSED APPROACH

There is a number of approaches, including various data mining and machine learning techniques, which could be applied for implementation of such user-based CF approach for content recommendation (Beel et al.[4], [5], Waila et al. [6], Sarwar et al. [7], Allen et al. [8]). In opting for a most suitable technique, it should be kept in mind that the OCW platform will eventually possess large, constantly growing and highly variable record store (so called User Activity Record Store) of user activities and preferences, as it will be constantly updated with new content and new users. Upon such User Activity Record Store, user-based CF algorithm will have to perform analysis in terms of similarity search and extraction of content to be recommended. Having this in mind, so called model-based CF recommendation (Xiaoyuan et al. [9]) that requires classical training or coefficients estimation of underlying model (such as Bayesian networks, clustering models, ANNs, SVMs, etc.) will not provide sufficiently generalization capabilities in terms of tackling new content and users without permanent (at least iterative) training and coefficients fitting. At the end, it will deliver large, computationally demanding, complex models (due to dimensionality of the problem) requiring permanent adjustments of underlying model. Answer to this problem can be provided by memory-based CF approach, which does not require any "old-fashion" model training and provides solution directly applicable upon the User Activity Record Store in finding most similar users.

Having previously mentioned in mind, as one of potential techniques suitable for implementation of memory-based CF approach would be k-Nearest Neighbours (k-NN) algorithm as one of the widely utilized techniques in machine learning and data mining. More precisely, k-NN algorithm represents a non-parametric method used for classification and regression. As such, k-NN algorithm can be used for implementation of (memory-based) user-based CF content recommendation. k-NN algorithm will perform the similarity search upon the User Activity Record Store to find $k$ most similar users, based on the user-related features (user activity and preferences). Each user represents a point in multidimensional space of user features (visited decks, liking & rating, commenting, etc.), as shown in Figure 1, whereby each user feature represents a single dimension:

$$u = [u_1, u_2, u_3, \dots, u_m],$$

where $u$ indicates the user, and $u_i$ indicates the user-related feature (in $m$-dimensional space where set of $m$ features define a user).

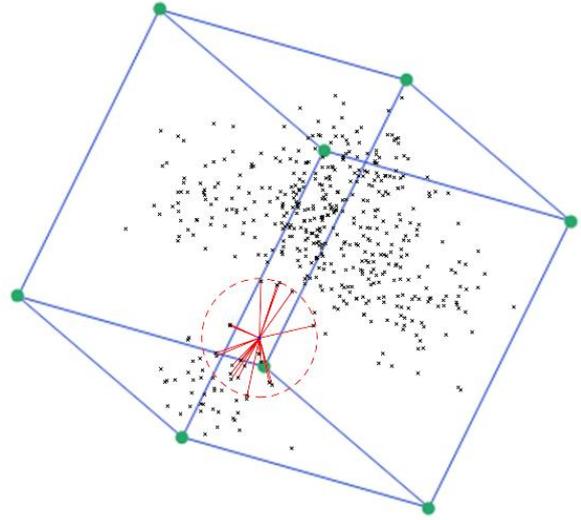

Figure 1. User's feature space (k-NN search algorithm).

In such multidimensional space, distance from the analysed user (the user whom the prediction is for) to other users will be calculated and $k$ nearest "neighbours" (i.e. the users with most similar past preferences and activities) will be extracted. User distance will be calculated based on Hamming distance (or Euclidian distance in case of real value features such as deck rates) as follows:

$$d_{u_i} = \sum_{i=1}^{m} w_i |u_i - u'_i|, \qquad u_i \in U,$$

where $d_{ui}$ indicates the distance between user $u_i$ and analysed user $u_i'$, $w_i$ weighting coefficient per user-related feature, while $u_i$ is taken from $U$ representing all users of the OCW platform. In order to discover $k$ nearest neighbours, corresponding weighting of user-related features (indicated by $w_i$) will be applied according to the feature relevance and potential impact on the recommendation decision. A particularly popular approach to the weights estimation in the literature is the use of evolutionary algorithms (e.g. genetic algorithm) to optimize the feature scaling (Nigsch et al. [10]). Finally, by cross-matching of the OCW content that was associated to these $k$ nearest neighbours (according to visits, comments, positive rates, etc.), top $n$ relevant OCW decks (that were not already associated to the user whom the prediction is for) will be recommended.

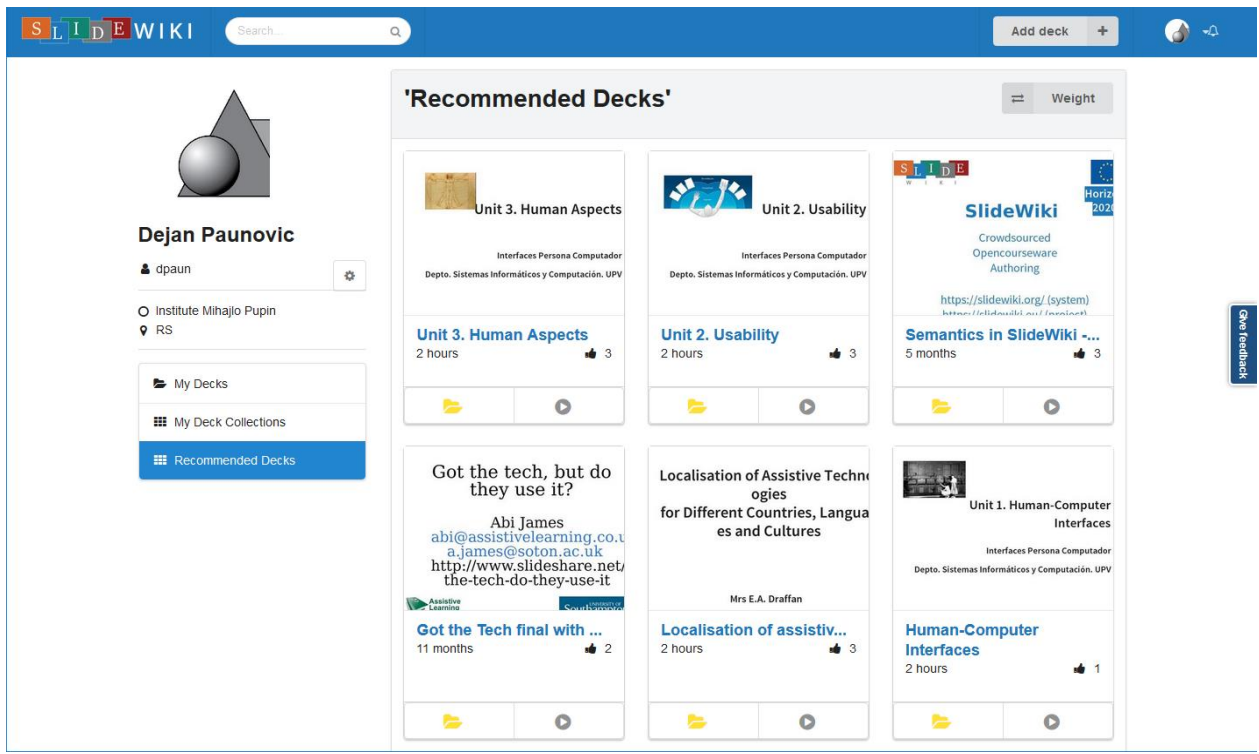

Figure 2. Recommender User Interface (SlideWiki platform).

III. USER-RELATED FEATURES

User-based CF recommendation module, leveraged upon k-NN algorithm, will analyse OCW users based on predefined set of relevant user-related features. As previously mentioned, this set of relevant user-related features will unambiguously represent each user as a specific point in multidimensional vector space. As part of relevant user-related features recommender will take the following:

- User activities (e.g. duration and number of deck visits, deck creation/editing, search activity, participation to related discussion, deck commenting, etc.),
- User preferences (e.g. likes and ratings of decks), and
- User demographic & contextual data (e.g. age, skills, date and time, location, etc.).

All data, i.e. relevant user-related features, should be collected implicitly and automatically by OCW platform (after appropriate anonymization of user data) in User Activity Record Store upon which k-NN algorithm will perform similarity search in order to discover content for recommendation. Moreover, to provide as rich database for post-processing as possible, additional explicit forms of data collection will be considered (e.g. asking a user to rate a content).

To provide efficient and precise content recommendation, apart from user activities and preferences, user demographic and contextual data will be also considered for processing as such data could reveal relevant behavioural patterns of OCW user that could have significant impact on content recommendation (so called context-aware collaborative filtering as specified by Adomavicius et al. [11]).

IV. RECOMMENDER IMPLEMENTATION

In order to evaluate and test the proposed user-based CF approach in a real content management environment, a designated content recommendation module leveraged by k-NN technique was implemented as part of the SlideWiki authoring OCW platform. SlideWiki platform is being developed as a part of the H2020 EU research and innovation project SlideWiki (Grant Agreement No 688095) aiming to create a large scale accessible learning and teaching platform, using educational technology, skill recognition and global collaboration. SlideWiki can be seen as a virtual learning environment, in which content editors and content owners can create new learning objects, generating knowledge and competences based in student's needs. Concretely, in SlideWiki the learning objects are represented as decks of slides, which are accessible, can be reused and allow student interactions. As Educational Recommender Systems (ERS) [13],[14] serve to recommend learning objects (decks in this case), according to the student's characteristics, preferences and learning needs, SlideWiki was aimed to analyze the profile of users considering user preferences, and topics and content details of the decks in which they show interest. Regarding content editors and content owners, it is possible also to establish their preferences based on their previous creations. In this regard, the proposed user-based CF approach using k-NN technique was implemented as part of the SlideWiki platform to serve as dedicated ERS. From the user deck page, users can access a list of recommended decks, which are computed by the platform using the proposed approach. Figure 2 shows a capture of this feature, in which the user can see a list of recommended decks, which are ordered by relevance (weights) from more interesting to less interesting for the user.

To validate the proposed approach, comprehensive validation process will be undertaken by tracking the preferences that user made and evaluating how they match with the recommended content. For validation purposes, different evaluation metrics will be taken into account and accommodated to reflect the efficiency of the proposed approach. Some of the commonly used metrics are the mean squared error and root mean squared error, while the information retrieval metrics such as precision and recall or discounted cumulative gain are useful to assess the quality of a recommendation method. Recently, diversity, novelty, and coverage are also considered as important aspects in evaluation process according to Lathia et al. [15]

## V. CONCLUSION

In this paper the conceptual idea of the content recommendation module was provided, which is capable of proposing the related content to the user having in mind past user activities, preferences, type and content similarity, etc. Suitable techniques for implementation of the user-based CF approach were analysed, and user-related features that are relevant for the content evaluation were identified. The proposed approach envisaged a hybrid recommendation system as a combination of user-based and content-based approaches in order to provide a holistic and efficient solution for content recommendation.

For evaluation and testing purposes, the proposed approach was implemented as a designated content recommendation module as part of the SlideWiki authoring OCW platform. In this regard, user-based CF recommender was deployed with dedicated user interface, where users can discover new content based on the preferences of other similar users. In this way, the proposed approach can help SlideWiki users to discover new content, so learners can access new decks to improve their skills, while content editors can check other similar decks to reuse their contents.

During the implementation and evaluation of content recommendation module, three well-known problems were identified that are specific to the collaborative filtering approaches (Lee et al.) [16]:

- *Cold start*: CF recommenders require a substantial amount of existing data on a user in order to make accurate recommendations (Rubens et al. [17], Elahi et al. [18]). Therefore, it is necessary to take into account the "initialization" phase for acquisition of user-related data before unlocking a full-blown potential of CF approach.
- *Scalability*: Environments in which CF should provide recommendations are often characterized with large amount of users and content, requiring high computation power. There are several approaches that could improve the scalability: k-NN search algorithm (optimised linear search; bottom up search - analysing per deck related activities and not per each activity solely), feature extraction & dimension reduction (clustering of similar decks/users; temporal constraint - only most recent activity is analysed), dataset partitioning (similarity search performed in parallel)
- *Sparsity*: As content database could be extremely large, often only the most active users are sharing their preferences (e.g. rating the content) just on a small subset of the overall content database. One of the remedies for this issue is to provide an explicit form of user preferences collection (e.g. asking a user to rate, search, to rank, create a list of favourite items, etc.) and generally to evoke the user activity on the platform.

As part of the future work and planned activities, each of these problems and related research challenges will be tackled and verified within the SlideWiki OCW platform to provide more efficient and reliable content recommendation system.

ACKNOWLEDGMENT

The research presented in this paper is partly financed by the European Union (H2020 SlideWiki project, Pr. No: 688095), and partly by the Ministry of Science and Technological Development of Republic of Serbia (SOFIA project, Pr. No: TR-32010).